\documentclass{article}
\usepackage{ijcai20}

\usepackage{times}
\usepackage{soul}
\usepackage{url}
\usepackage[hidelinks]{hyperref}
\usepackage[utf8]{inputenc}
\usepackage[small]{caption}
\usepackage{graphicx}
\usepackage{amsmath}
\usepackage{amsthm}
\usepackage{booktabs}
\usepackage{algorithm}
\usepackage{algorithmic}
\title{Keep It Real: a Window to Real Reality in Virtual Reality}

\author{
Baihan Lin
\affiliations
Computational Visual Inference Lab, Columbia University, New York, NY 10027, USA\\
Department of Applied Mathematics, University of Washington, Seattle, WA 98195, USA\\
\emails
baihan.lin@columbia.edu
}

\begin{document}

\maketitle

\begin{abstract}

This paper proposed a new interaction paradigm in the virtual reality (VR) environments, which consists of a virtual mirror or window projected onto a virtual surface, representing the correct perspective geometry of a mirror or window reflecting the real world. This technique can be applied to various videos, live streaming apps, augmented and virtual reality settings to provide an interactive and immersive user experience. To support such a perspective-accurate representation, we implemented computer vision algorithms for feature detection and correspondence matching. To constrain the solutions, we incorporated an automatically tuning scaling factor upon the homography transform matrix such that each image frame follows a smooth transition with the user in sight. The system is a real-time rendering framework where users can engage their real-life presence with the virtual space. \footnote{The data and code to reproduce the video application can be downloaded at \url{https://github.com/doerlbh/V2R}}
\end{abstract}

\section{Introduction}

Traditional VR technologies usually emphasize creating an entirely immersive environment, visual inputs of the real realities are usually discouraged \cite{burdea2003virtual,bowman2007virtual}. Studies have attempted to introduce virtual mirrors to facilitate embodiment \cite{spanlang2014build}, rehabilitation \cite{sato2010nonimmersive} and education \cite{blum2012mirracle}, but almost all of them reflects the virtual space (e.g. the body of the avatar in the virtual room). We demonstrated in our system, however, that the introduction of such a \textit{virtual} window or mirror reflecting the \textit{real} world can bridge the two realities in a surprisingly beneficial synergy: (1) introducing a virtual window or mirror which the user could activate or deactivate any time could encourage the user to actively attend to and seamlessly receive real-world feedbacks where virtual rendering was challenging; (2) projecting a geometrically accurate representation of users' image and reality into a movie could create an illusion that the user is presently sharing the same universe with the virtual characters -- breaking the fourth wall. 

\section{Projective Geometry}

\begin{figure}[tb]
\centering
    \includegraphics[width=\linewidth]{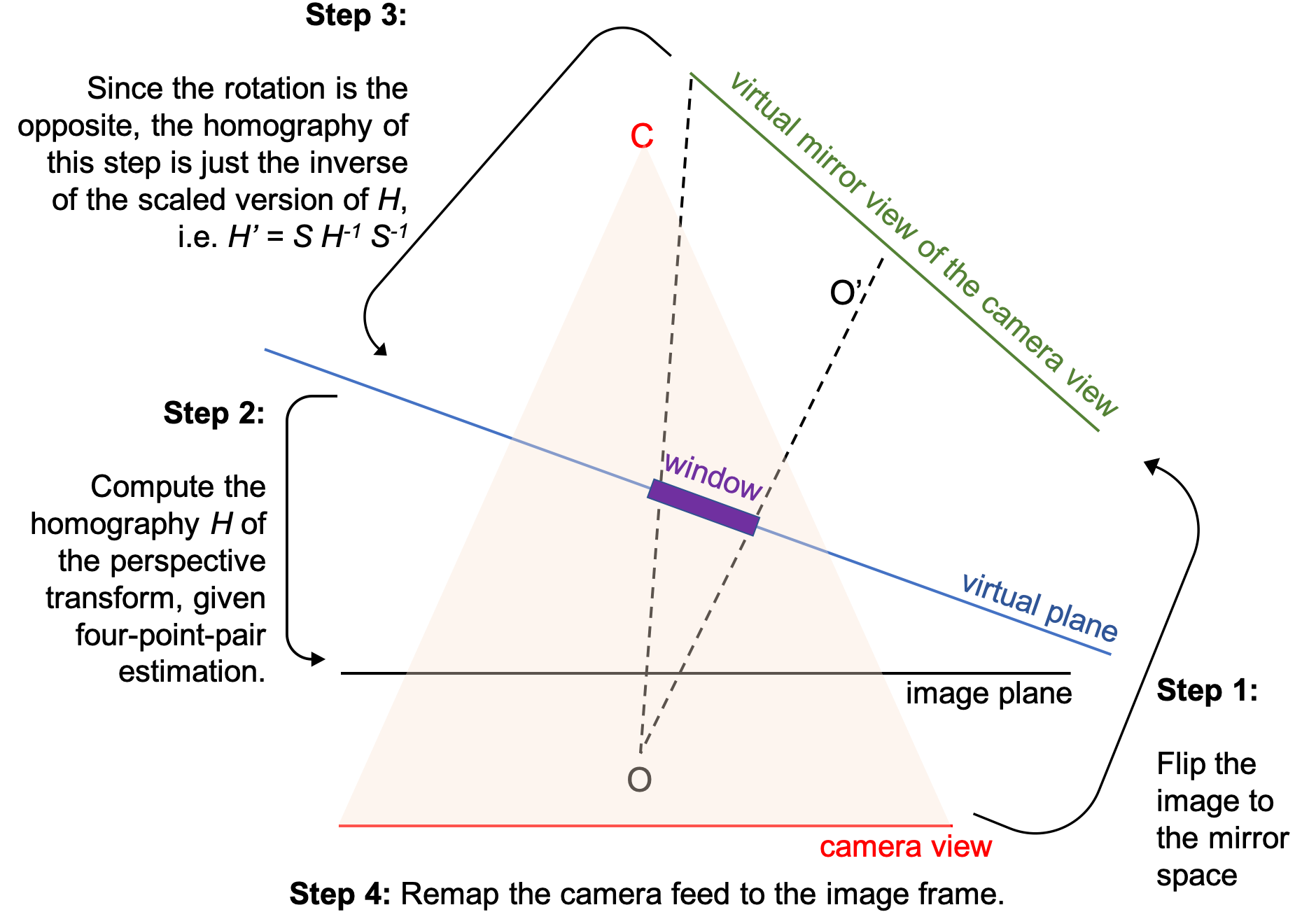}
    \par\caption{Projective geometry problem for the perspective mirror.}\label{fig:geometry}
\end{figure}

As in Figure \ref{fig:geometry}, the projective geometry problem is outlined as follows: the user is the observer $O$ who has a reflection of a virtual mirror image at $O'$ from a window or mirror on a virtual plane (e.g. a virtual wall); camera $C$ with unknown distances covers the user in its field angle, and we wish to solve the perspective transform (the homography), from the camera view plane, (i) to the virtual camera view plane by the mirror, (ii) to the virtual plane where the mirror is located, and eventually (iii) to the image space where the observer sees. Solving (iii) is straightforward \cite{coxeter2003projective}, but solving (i) and (iii) is nontrivial since depth estimation in a video and the camera at the same time is an innately chanllenging problem. In another word, there is not a unique homography for the projection of the mirror image, but a ray of solutions for each point. We tackled the issue by introducing an additional scaling factor to constrain the homography according to each pairs of image frame. The scheme follows Figure \ref{fig:geometry}.

\begin{figure}[tb]
\centering
    \includegraphics[width=0.24\linewidth]{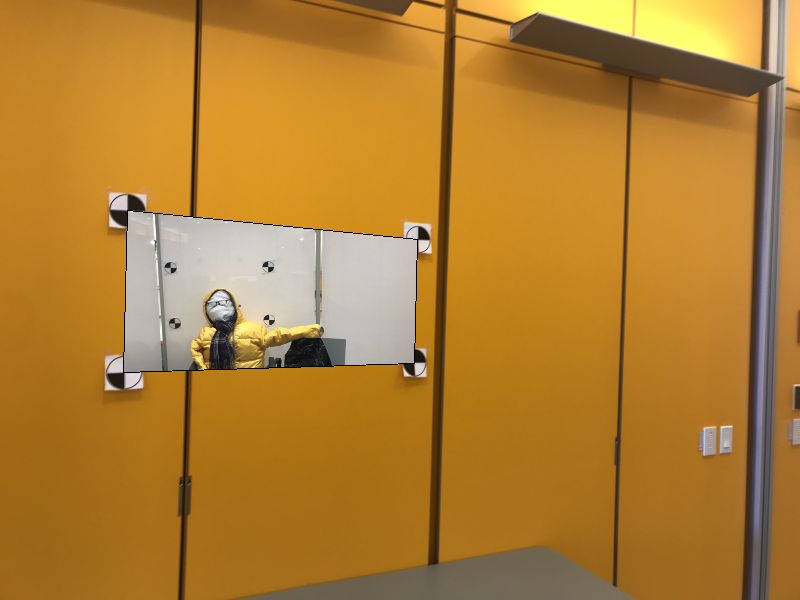}
    \includegraphics[width=0.24\linewidth]{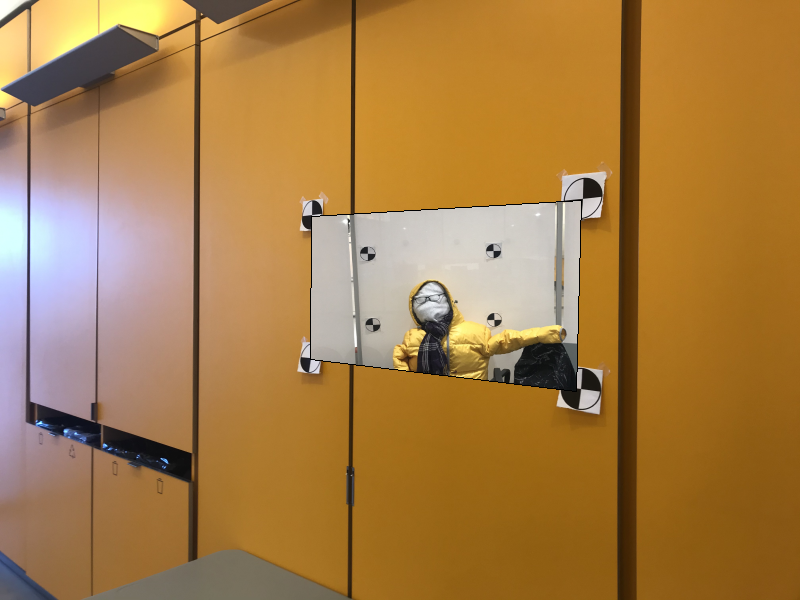}
    \includegraphics[width=0.24\linewidth]{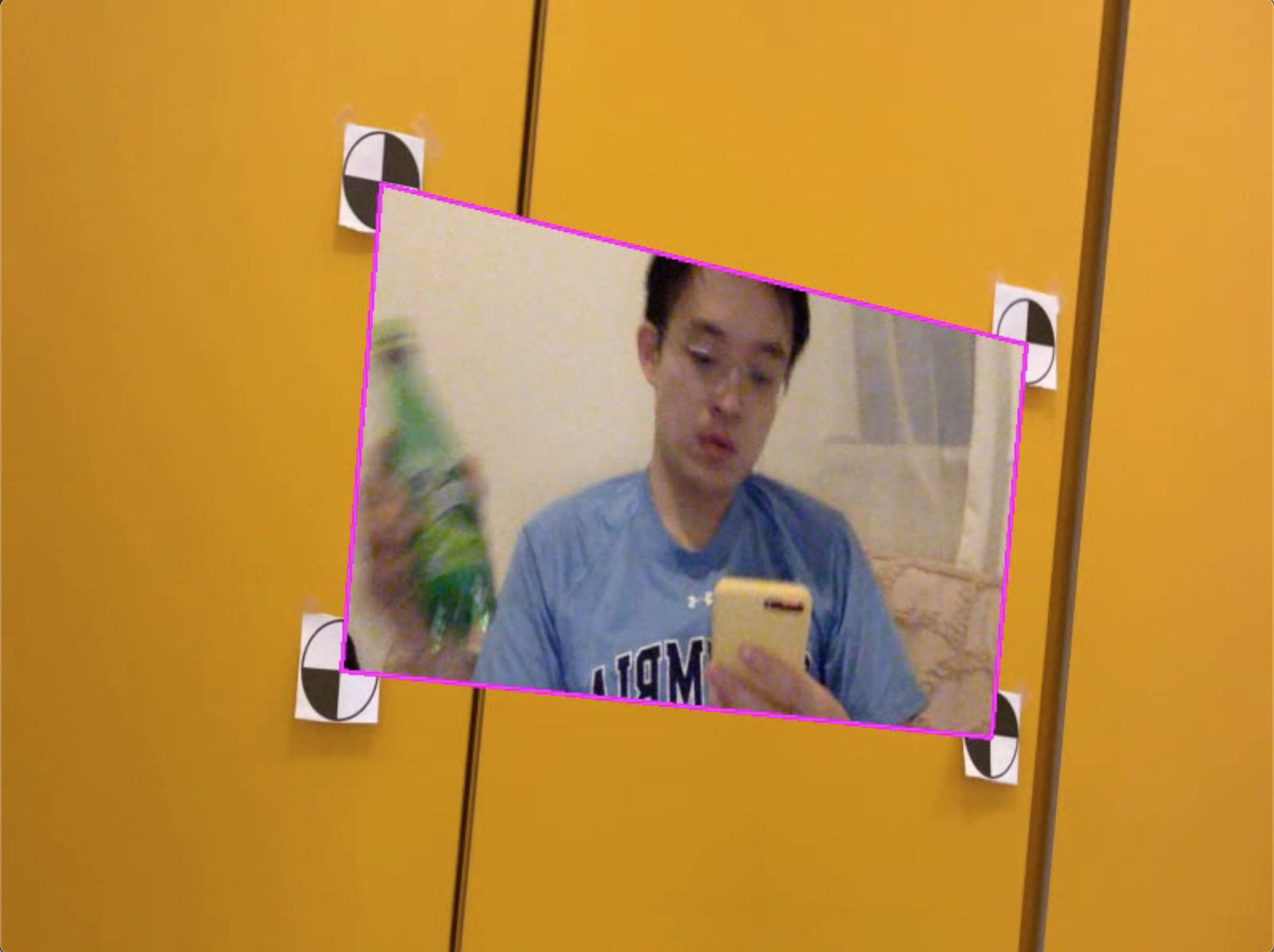}
    \includegraphics[width=0.24\linewidth]{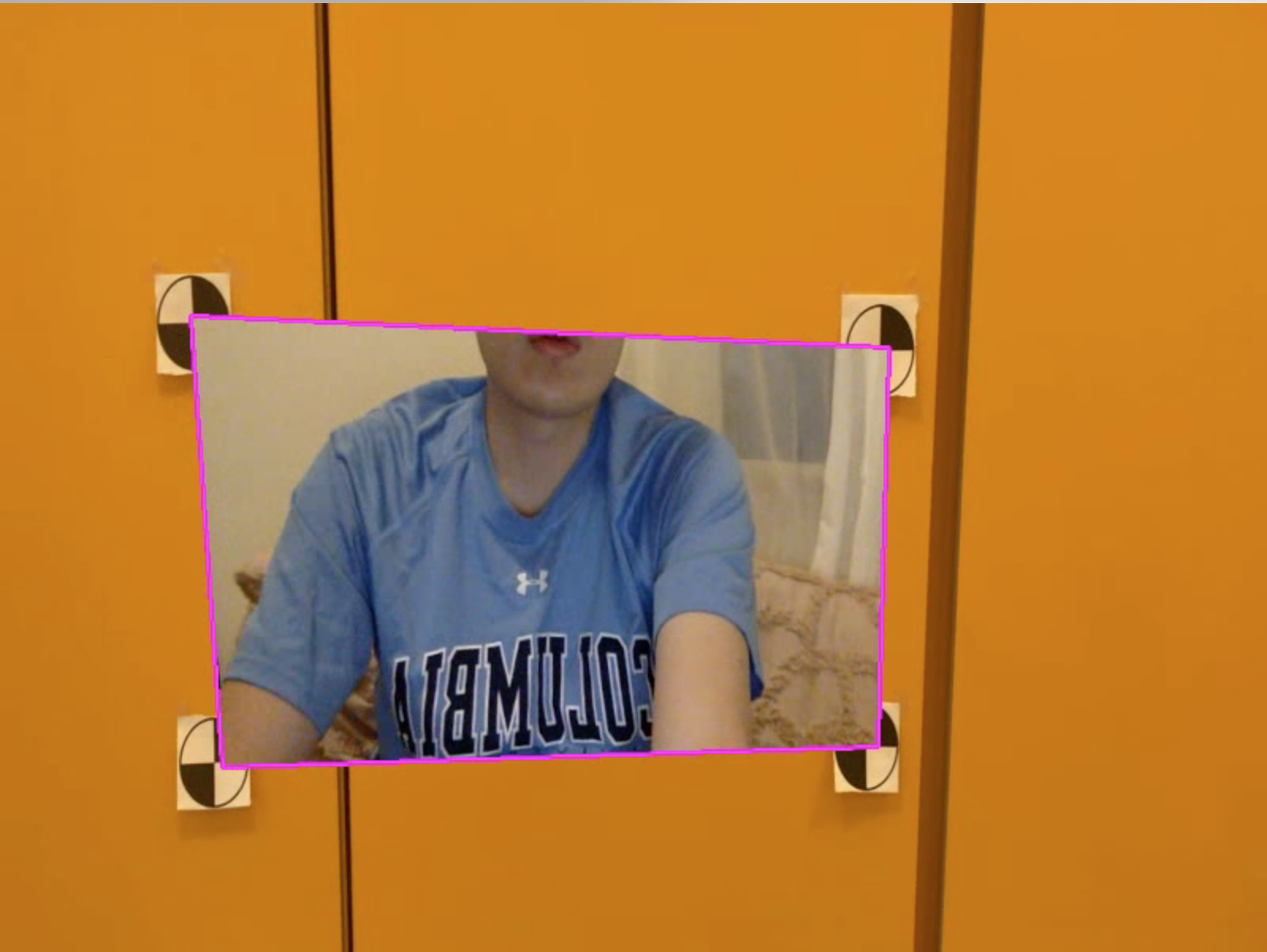}
\par\caption{Examples of the perspective mirrors in augmented reality.}\label{fig:results}
\end{figure}

\begin{figure*}[t]
\centering
    \includegraphics[width=0.9\linewidth]{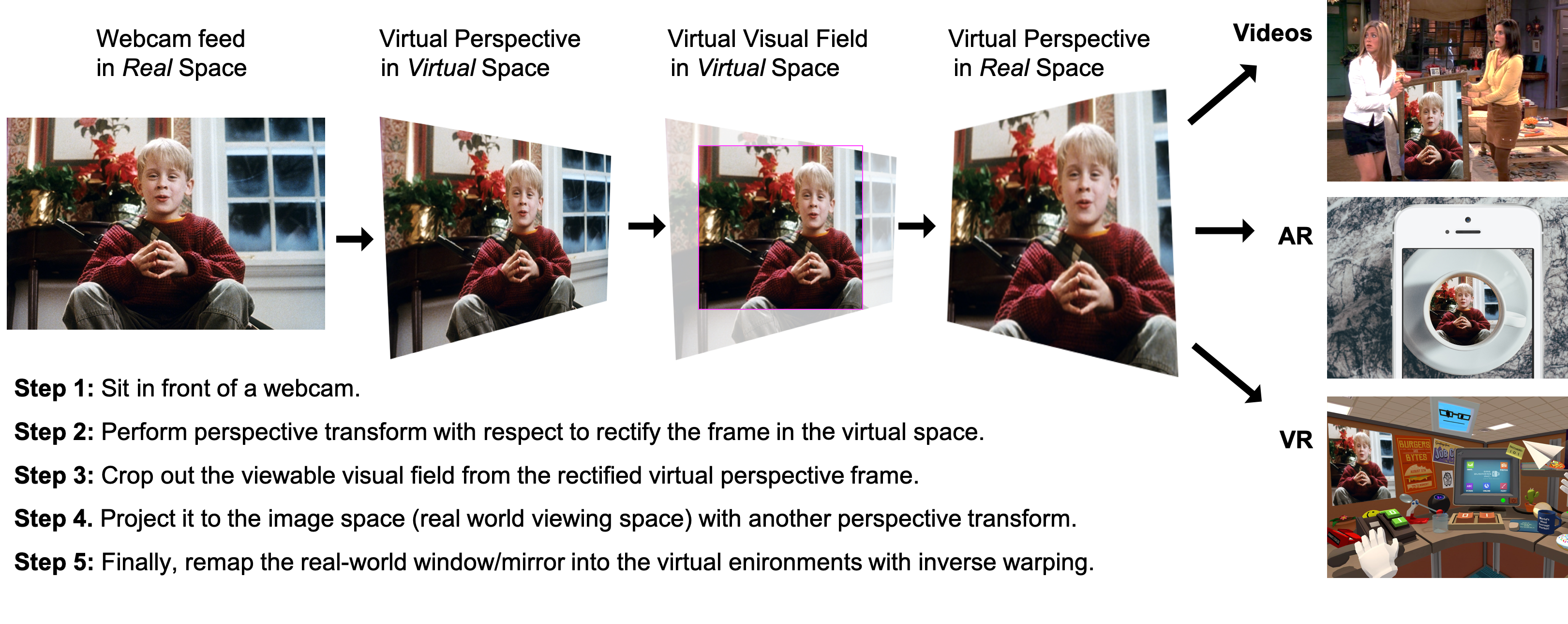}
    \vspace{-1em}
    \caption{Procedures to create a real-world window or perspective mirror in the virtual environments (videos, stream apps, AR and VR).}\label{fig:mirror}
\end{figure*}

\section{Virtual-to-Real (V2R) Mirrors or Windows}

The V2R framework involves two main components: pattern matching (where to put our windows) and image warping. 

The pattern matching problem is defined as a feature detection plus classification problem (in case of multiple windows or mirrors). Since we would like to apply it to real-time camera feed (as in AR and VR), there are two major difficulties: (1) a short response time requires the model to slim and fast; (2) noisy, blurring and shaking image frame in the video due to hand-held motion. Neural networks tend to be slower so we resort to the signal processing methods (such as filtering) to speed up the computation. For demonstration purpose, we chose the classical cross-shaped black-white circle as markers to benchmark our methods, and obtained robust performance with template matching, Harris and Hough operators \cite{brunelli2009template} after applying Guassian and median filters. 

Figure \ref{fig:mirror} summarizes the steps involved in the image warping: first, rectify the camera feed with respect to the virtual space; then, crop the image given a proper camera field angle; and finally project it the image space given the computed homography. This framework turns out to be fairly robust and flexible to apply to a wide range of applications. For instance, we can introduce a V2R mirror into a movie in every scene where a target painting appears, as in the example of the TV series ``Friends''. In VR, we can also make the postboard a V2R window whenever a button on the control bar is pressed.

Figure \ref{fig:results} confirmed the validity of our method: for instance, in the rightmost case, the virtual perspective is clearly looking down from above, and the warped image is indeed as if the observer was looking down and can only see from the mirror the bottom half of the body. V2R can be applied in real-time in a timely fashion (more example videos and images can accessed at \url{https://github.com/doerlbh/V2R}). 

As an important concept of AI, the sense of presence can effectively engage the interaction between the human with the virtual agents in virtual environments. V2R aims to create such a synergy by bridging the two spaces seamlessly.  

\section{The Demonstration System}

V2R provides extensive entertainment capacities and interactive possibilities, such as activating and deactivating the window given certain features or user commands (in VR), experiencing the thrilling emotion of being in the same room with the serial killer in a blockbuster horror movie only a mirror reflection away (in movies), blending your face with an ancient guru master through the secret portal of a crystal ball (in AR), priming customers to buy a product by projecting users' real world objects into a commercial (in online streaming ad). In the demonstration, we will provide a pool of interesting short videos, where the mirror location detection have been properly trained and tuned, such that the user can see themselves instantly within these movie clips (e.g. as in Figure \ref{fig:results}).

\section{Ongoing and Future Directions}

Exciting next steps involve developing a wide spectrum of interactions around V2R and accompany intelligent modules such as gesture recognition (imagine creating a virtual window to the real world by drawing a square in the air with your hands), video conference with virtual windows to multiple camera feeds (literally as if in the same room), real-time rendering of virtual objects given the real-world camera feed (bring an apple from the real life to the virtual space) etc.. Another direction is to improve the projective geometry: these mirrors and windows projected from single-camera 2d images are currently pseudo-representations of what one should see from the exact computed point of view in the real world, as the single-camera setup make it hard to perform a 3d reconstruction. A useful extension of V2R is to introduce multiple cameras, such that the camera depth can be properly estimated along with a 3d projection of the details that could be revealed but hidden in 2d representation (e.g. an occlusion). Another useful extension is to introduce visuo-tactile and visuo-motor feedbacks \cite{berger2020follow} into this blended setting, which can potentially change peripersonal space in VR. These improvements should illicit highly interactive user experience and engage more users and research into the field of intelligent virtual reality interactions.

\bibliographystyle{named}
\bibliography{main}

\begin{thebibliography}{}

\bibitem[\protect\citeauthoryear{Berger \bgroup \em et al.\egroup
  }{2020}]{berger2020follow}
Christopher Berger, Baihan Lin, Bigna Lenggenhager, Jaron Lanier, and Mar
  Gonzalez-Franco.
\newblock {Follow Your Nose: Extended Arm Reach After Pinocchio Illusion in
  Virtual Reality}.
\newblock {\em under review}, 2020.

\bibitem[\protect\citeauthoryear{Blum \bgroup \em et al.\egroup
  }{2012}]{blum2012mirracle}
Tobias Blum, Valerie Kleeberger, Christoph Bichlmeier, and Nassir Navab.
\newblock mirracle: An augmented reality magic mirror system for anatomy
  education.
\newblock In {\em 2012 IEEE Virtual Reality Workshops (VRW)}, pages 115--116.
  IEEE, 2012.

\bibitem[\protect\citeauthoryear{Bowman and McMahan}{2007}]{bowman2007virtual}
Doug~A Bowman and Ryan~P McMahan.
\newblock Virtual reality: how much immersion is enough?
\newblock {\em Computer}, 40(7):36--43, 2007.

\bibitem[\protect\citeauthoryear{Brunelli}{2009}]{brunelli2009template}
Roberto Brunelli.
\newblock {\em Template matching techniques in computer vision: theory and
  practice}.
\newblock John Wiley \& Sons, 2009.

\bibitem[\protect\citeauthoryear{Burdea and Coiffet}{2003}]{burdea2003virtual}
Grigore~C Burdea and Philippe Coiffet.
\newblock {\em Virtual reality technology}.
\newblock John Wiley \& Sons, 2003.

\bibitem[\protect\citeauthoryear{Coxeter}{2003}]{coxeter2003projective}
Harold Scott~Macdonald Coxeter.
\newblock {\em Projective geometry}.
\newblock Springer Science \& Business Media, 2003.

\bibitem[\protect\citeauthoryear{Sato \bgroup \em et al.\egroup
  }{2010}]{sato2010nonimmersive}
Kenji Sato, Satoshi Fukumori, Takashi Matsusaki, Tomoko Maruo, Shinichi
  Ishikawa, Hiroyuki Nishie, Ken Takata, Hiroaki Mizuhara, Satoshi Mizobuchi,
  Hideki Nakatsuka, et~al.
\newblock Nonimmersive virtual reality mirror visual feedback therapy and its
  application for the treatment of complex regional pain syndrome: an
  open-label pilot study.
\newblock {\em Pain medicine}, 11(4):622--629, 2010.

\bibitem[\protect\citeauthoryear{Spanlang \bgroup \em et al.\egroup
  }{2014}]{spanlang2014build}
Bernhard Spanlang, Jean-Marie Normand, David Borland, Konstantina Kilteni,
  Elias Giannopoulos, Ausi{\`a}s Pom{\'e}s, Mar Gonz{\'a}lez-Franco, Daniel
  Perez-Marcos, Jorge Arroyo-Palacios, Xavi~Navarro Muncunill, et~al.
\newblock How to build an embodiment lab: achieving body representation
  illusions in virtual reality.
\newblock {\em Frontiers in Robotics and AI}, 1:9, 2014.

\end{thebibliography}

\end{document}